\documentclass[a4paper,11pt]{article} 
\usepackage{graphicx}
\usepackage{epsfig}
\begin{document}
\begin{center}
{\large \bf \sf 
 \bf 
INSTABILITIES OF MHD WAVES PRODUCED BY COUPLING OF ROTATION AND GRADIENT OF MAGNETIC
FIELD AND ITS POSSIBLE APPLICATION IN THE GALACTIC CENTRAL REGION}\\
\vskip 1 cm
{\bf Ipsita Das} \footnote{e-mail: ipsitadas23@gmail.com} and 
{\bf  Ramaprosad Bondyopadhaya} \footnote{e-mail: rpb\_ jumath@email.com}\\
\vskip 0.25 cm
{\em Department of Mathematics, Jadavpur University, Kolkata-700032,India .}\\
\end{center} 
\vskip 2.5 cm 
{\bf Abstract : }\\

An analysis of MHD wave propagating in a gravitating and rotating medium
permeated by non-uniform magnetic field has been done. It has been found 
that the Gradient of Magnetic Field when coupled with Rotation becomes 
capable to generate few instabilities (Temporal or Spatial) leading to 
the damping or amplification
of MHD waves. The Jean's criterion is not sufficient for stability always. 
Rather, the waves will suffer instability unless their wave length
(frequency) is less (greater) than certain critical values. 
Otherwise, those will smoothly propagate outward. 
Out of different scenarioes depending on the direction of the magnetic field, 
its gradient, rotation and wave propagation three important Special Cases have 
been discussed and different stability criteria have been derived.\\

Finally, using the above theory we have obtained the stability/instability 
criteria for the waves moving 
parallel and perpendicular to the galactic plane in the Core and Periphery
of the Central Region of Galaxy (C.R.G.) {\bf due to the coupled action of 
Rotation and Non-Uniform Magnetic field.} 
The possibility of heating or occuring diffused condition inside the 
central region by MHD waves or smooth propagation of these waves (under 
some restrictions) through the C.R.G. has been briefly discussed.
The numerical values of the parameters of those waves for instabilities or smooth
propagation have been estimated roughly.
One may find some clues for the formation of Halo and Spiral Arms.\\ 

{\bf keywords:}
 MHD Instability, Magnetic field gradient and rotation, Central region of galaxy.
\newpage
\begin{center}
{\bf 1. INTRODUCTION}
 \end{center} 
 Any disturbance propagating through a non-uniform MHD fluid or plasma may be stable or undergoes 
instability. The damping of MHD waves 
is one of the causes of heating of such medium (Sturrock, 1966; Barnes, 1968; Bondyopadhaya et. al. 
1972, 1974). In general, if streaming flow is present it may help instability in this medium 
(Sturrock, 1960; Gold, 1965; Sudan, 1965). The strong rotation perpendicular to gravity, however, 
may help the stability of magnetized fluid (Gilman, 1970). The rotation can influence in the 
damping or heating of MHD medium like stellar interior (Bondyopadhaya 1972, 1974, 1978). The 
MHD instability due to temperature gradient and possibility of mass-outflow from the central
region of galaxy has been investigated by Chakraborty and Bondyopadhaya (1998), Chakraborty 
(1998, Thesis). Sarkar and Bondyopadhaya (2007) have discussed the MHD instability due to non-uniform 
magnetic field and its effect on the propagation of waves of wave length greater 
than certain critical value. They also discussed the role of MHD instability in heating due to damping of waves having 
length less than that value. In reality, MHD flow seems to be connected with different physical 
processes of AGN (See, e.g. Blandford, 1990) or MHD Jets from AGN (See, e.g. Rosen and Hardee, 2000; Wiita, 2002).

In this paper we have considered a gravitating and rotating MHD medium permeated by non-uniform
magnetic field. We shall use the perturbation technique to linearise the guiding equations.
Assuming the wave structure of the perturbation, General Dispersion Relation will be derived. Next
investigation will be made about the wave propagation in different situations characterised 
by the Magnetic Field 
Gradient and Rotation. A number of stability/instability criteria will be derived which, 
specialy,
are produced by the coupled action of Non-Uniform Magnetic Field and Rotation. 
Finally, these results will be used to have better understanding of instabilities (heating or 
diffused condition or other phenomena) or stabilities (smooth wave propagation)occuring in Galactic
Central Region.

\begin{center}
\bf 2. BASIC EQUATIONS
 \end{center}
Let us consider a gravitating, rotating MHD fluid medium having finite conductivity. The basic equations of this
 fluid flow are the following :---- \\
\\
$\partial \vec u / \partial t= 1 / (4\pi \mu \rho) (\nabla \times  \vec B)\times \vec B - (\nabla \times  \vec u)
 \times \vec u - \rho^{-1}   \nabla p$ \\$~~~~~~~~~~~~~-\nabla ( u ^{2}/2 + \phi ) - 2(\vec w \times \vec u )
~~~~~~~$[Equation of Motion]~~~~~~~~
~~~~~~~~~~~~~~~~~~~~~~~~~~~~~~(1) \\
$\partial \vec B/ \partial t = \nabla \times (\vec u \times \vec B) + \eta \nabla ^{2}\vec B$~~~~~~~~~
~~~~~~[MHD Field Equation]
~~~~~~~~~~~~~~~~~~~~~~~~~~~~~~~~~~~~~~~~~(2)\\
$\partial \rho /\partial t = -\nabla (\rho \vec u)$~~~~~~~~~~~~~~~~~~~~~~~~~~~~~~~~~~~[Continuity Equation]
~~~~~~~~~~~~~~~~~~~~~~~~~~~~~~~~~~~~~~~~~ (3)\\ 
$\nabla ^{2} \phi = 4\pi G\rho $~~~~~~~~~~~~~~~~~~~~~~~~~~~~~~~~~~~~~~~[Poisson's Equation]
~~~~~~~~~~~~~~~~~~~~~~~~~~~~~~~~~~~~~~~~~(4)\\
$p = (C_{s}^{2} /\gamma )\rho $~~~~~~~~~~~~~~~~~~~~~~~~~~~~~~~~~~~~~~~~~ [Equation of State] 
~~~~~~~~~~~~~~~~~~~~~~~~~~~~~~~~~~~~~~~~~~~~~~(5)\\
\\
 where, \\
 $\vec u$ = Fluid velocity~~~~~~~~~~~~~~~~~~~~~~~~~~~$\vec w$ = Angular velocity\\
 $\vec B$ = Magnetic field~~~~~~~~~~~~~~~~~~~~~~~~~~$p$ = Hydrostatic pressure\\
 $\rho $ = Gas density~~~~~~~~~~~~~~~~~~~~~~~~~~~~~~$G$ = Gravitational constant\\                                     
 $\phi $ = Gravitational potential~~~~~~~~~~~~~~ $\eta = c/4\pi \mu \sigma $ = Electrical Resistivity\\
 $\mu $ = Permeability~~~~~~~~~~~~~~~~~~~~~~~~~~~~~$\sigma $ = Electrical Conductivity\\
 $C_{s} $ = Sound speed~~~~~~~~~~~~~~~~~~~~~~~~~~~~$c$=Velocity of light in vaccum\\
 $\gamma $=Ratio of specific heats\\
 All other symbols have there usual meaning.
 
 \begin{center}
\bf 3. PERTURBED LINEARISED EQUATION\\
 \end{center} 
  We shall now consider the situation when initially all the variables was at equilibrium 
but subsequently made perturbed as $\psi = \psi _{0} + \psi \prime $ where $\psi _{0}$ 
and $\psi \prime $ are respectively unchanged value and small perturbation value of the 
variable $\psi$ . First all the variables of equations (1) to (5) are perturbed, then the equilibrium 
conditions (obtained by putting equilibrium value in (1) to (5)) are used and finally the square or higher power 
of perturbation variables are neglected so that we get the following linearised equations 
( assuming, however, there is no initial streamflow i.e. $u_{0} = 0$ and $\gamma =1$ )
(See, Sarkar and Bondyopadhaya, 2007)\\ 
\\
$ \partial \vec u/\partial t = \alpha[(\nabla \times \vec B)\times \vec B_{0} + 
(\nabla \times \vec B_{0})\times \vec B]- (\rho _{0}^{-1}) \nabla p - \nabla \phi  - 2(\vec w_{0} \times \vec u)
~~~~~~~~~~~~~~~~~~~~~~~~$(6) \\
$ \partial \vec B/\partial t = \nabla \times (\vec u \times \vec B_{0}) + \eta \nabla ^{2}\vec B ~~~~
~~~~~~~~~~~~~~~~~~~~~~~~~~~~~~~~~~~~~~~~~~~~~~~~~~$~~~~~~~~~~~~~~~~~~~~~~~~~~~~(7) \\
$ \partial \rho /\partial t = -\rho _{0} \nabla \cdot \vec u $ ~~~~~~~~~~~~~~~~~~~~~~~~~~~~~~~~~~~~~~~~~~~~~~~~~~~~~~
~~~~~~~~~~~~~~~~~~~~~~~~~~~~~~~~~~~~~~~~~~~~~~~~~~(8) \\
$ \nabla ^{2} \phi = 4\pi G\rho $ ~~~~~~~~~~~~~~~~~~~~~~~~~~~~~~~~~~~~~~~~~~~~~~~~~~~~~~~~~~~~~~~~~~~~
~~~~~~~~~~~~~~~~~~~~~~~~~~~~~~~~~~~~~~~~~~~~(9) \\
$ p = C_{s}^{2} \rho $ ~~~~~~~~~~~~~~~~~~~~~~~~~~~~~~~~~~~~~~~~~~~~~~~~~~~~~~~~~~~~~~~~~~~~~~~~~~~~~~~~~
~~~~~~~~~~~~~~~~~~~~~~~~~(10) \\

where\\
 $C_{s}^{2} = (R T_{0})$,~~~~~~$T_{0}$=Initial Temperature, ~~~~~~~~~~$\alpha =1/(4\pi \mu \rho _{0})$ \\
$R$= Gas constant(obtained by dividing Universal Gas constant by mean 
mol. wt. of the gas).\\
Here the primes have been droped and suffixes zero denote the equilibrium values.

\begin{center}
\bf 4. PERTURBATION STRUCTURE AND GUIDING EQUATION\\
 \end{center}
Let us assume that all the perturbation parameters are proportional to Exp i(kx-$\omega $t) i.e. 
they posses wave like structure , having the direction of propagation as x-axis, k and $\omega $ being 
the wave number and wave frequency respectively. Then using the equations (8),(9) and (10)
we can write the equations (6) and (7) componentwise as follows : \\

From Equation of Motion :\\
$u_{x} (\omega ^{2} -k^{2}c_{s}^{2}+w_{g}^{2})/\omega +u_{y}(-2iw_{0z})+u_{z}(2iw_{0y})+\\
B_{y}[\alpha (-kB_{0y}+i(\partial {B_{0y}}/\partial {x}-\partial {B_{0x}}/ \partial {y}))]+
B_{z}[\alpha (-kB_{0z}-i(\partial {B_{0x}}/\partial {z}-\partial B_{0z}/\partial {x}))]=0 $~~~~~~~~~~~~~(11)\\
\\
$u_{x} (2iw_{0z}) +u_{y}(\omega )+u_{z}(-2iw_{0x})+\\
B_{x}[-i\alpha (\partial {B_{0y}}/\partial {x}-\partial {B_{0x}}/ \partial {y})]+B_{y}(k\alpha B_{0x})+
B_{z}[i\alpha (\partial {B_{0z}}/\partial {y}-\partial B_{0y}/\partial {z})]=0~~~~~~~~~~~~$(12)\\
\\
$u_{x} (-2iw_{0y}) +u_{y}(2iw_{0x} )+u_{z}(\omega )+\\
B_{x}[i\alpha (\partial {B_{0x}}/\partial {z}-\partial {B_{0z}}/ \partial {x})]
+B_{y}[-i\alpha (\partial {B_{0z}}/\partial {y}-\partial B_{0y}/\partial {z})]
+B_{z}(k\alpha B_{0x})=0~~~~~~~~~~~~~~~~~~~~$(13)\\
\\
From Field Equation :\\
$u_{x}[-i(\partial B_{0y}/ \partial {y} +\partial B_{0z}/\partial {z} )]+u_{y}(i\partial B_{0x}/ \partial {y} )
+u_{z}(i\partial B_{0x}/\partial {z})+B_{x}(\omega +i\eta  k^{2})=0~~~~~~~~~~~~~~$(14)\\
\\
$u_{x}(-kB_{0y} +i\partial B_{0y}/\partial {x} )+u_{y}[kB_{0x}-i(\partial B_{0x}/ \partial {x}+
\partial B_{0z}/\partial {z} )]+u_{z}(i\partial B_{0y}/\partial {z})+\\
B_{y}(\omega +i\eta  k^{2})=0~~~~~~~~~~~~~~~~~~~~~~~~~~~~~~~~~~~~~~~~~~~~~~~~~~~~~~~~~~~~~~~~~~~~~~~~~~~
~~~~~~~~~~~~~~~~~~~~~~~~~~~~~~$(15)\\
\\
$u_{x}(-kB_{0z} +i\partial B_{0z}/\partial {x} )+u_{y}(i\partial B_{0z}/\partial {y})  +u_{z}
[kB_{0x}-i(\partial B_{0x}/ \partial {x}+\partial B_{0y}/\partial {y} )]+\\
B_{z}(\omega +i\eta  k^{2})=0$~~~~~~~~~~~~~~~~~~~~~~~~~~~~~~~~~~~~~~~~~~~~~~~~~~~~
~~~~~~~~~~~~~~~~~~~~~~~~~~~~~~~~~~~~~~~~~~~~~~~~~~~~~~~~~~(16)\\
\\
The equations can be put in matrix form as,\\
\\
AX=0~~~~~~~~~~~~~~~~~~~~~~~~~~~~~~~~~~~~~~~~~~~~~~~~~~~~~~~~~~~~~~~~~~~~~~~~~~~~~~~~~~~~~~~~~~~~~~~~
~~~~~~~~~~~~~~~~~~~~ (17)\\
\\
where, X=$(u_{x}~~ u_{y}~~ u_{z}~~ B_{x}~~ B_{y}~~ B_{z})^{T}$  is a column matrix and 
$A=(a_{ij})$ is a square matrix of order 6, $a_{ij}$ denoting the coefficient of j-th variable in the 
i-th equation, i,j=1 to 6, has the following values: \\
\\
$a_{11}=(\omega ^{2}-k^{2}c_{s}^{2}+w_{g}^{2})/\omega, ~ a_{12}=-2iw_{0z}, ~ a_{13}=2iw_{0y},a_{14}=0,\\
a_{15}=\alpha [-kB_{0y}+i(\partial {B_{0y}}/\partial {x}-\partial {B_{0x}}/ \partial {y})], ~  
a_{16}=\alpha [-kB_{0z}-i(\partial {B_{0x}}/\partial {z}-\partial B_{0z}/\partial {x})] $\\
$  a_{21}=2iw_{0z}, ~  a_{22}=\omega , ~ a_{23}=-2iw_{0x}, ~ a_{24}=-i\alpha(\partial {B_{0y}}/\partial {x}-
\partial {B_{0x}}/ \partial {y}), ~ a_{25}=k\alpha B_{0x},\\
a_{26}=i\alpha (\partial {B_{0z}}/\partial {y}-\partial B_{0y}/\partial {z}) $\\
$a_{31}=-2iw_{0y}, ~ a_{32}=2iw_{0x}, ~ a_{33}=\omega , ~ a_{34}=i\alpha (\partial {B_{0x}}/\partial {z}-
\partial 
{B_{0z}}/ \partial {x}),\\
a_{35}= -i\alpha (\partial {B_{0z}}/\partial {y}-\partial B_{0y}/\partial {z}), a_{36}=k\alpha B_{0x}$\\
$a_{41}=-i(\partial B_{0y}/ \partial {y} +\partial B_{0z}/\partial {z} ), ~ a_{42}=i\partial B_{0x}/ 
\partial {y}, ~ a_{43}=i\partial B_{0x}/\partial {z}, ~ a_{44}=\omega +i\eta  k^{2},\\
a_{45}=0, a_{46}=0$\\
$a_{51}=-kB_{0y} +i\partial B_{0y}/\partial {x}, ~ a_{52}=kB_{0x}-i(\partial B_{0x}/ \partial {x}+
\partial B_{0z}
/\partial {z} ), ~ a_{53}=i\partial B_{0y}/\partial {z}, ~ a_{54}=0,\\
a_{55}=a_{44}, ~ a_{56}=0 $\\
$a_{61}=-kB_{0z} +i\partial B_{0z}/\partial {x}, ~ a_{62}=i\partial B_{0z}/\partial {y}, ~ a_{63}=kB_{0x}-
i(\partial B_{0x}/ \partial {x}+\partial B_{0y}/\partial {y} ), ~ a_{64}=0,\\
a_{65}=0, ~ a_{66}= a_{55}= a_{44} $.\\
\\
where, $w_g ^2=4\pi G \rho_0$ = Gravitational potential.

\begin{center}
\bf 5. GENERAL DISPERSION RELATION (G.D.R.)\\
 \end{center}
Eliminating the variables from the equation (17) we get the General Dispersion Relation (G.D.R.) as, \\
\\
det(A)=0~~~~~~~~~~~~~~~~~~~~~~~~~~~~~~~~~~~~~~~~~~~~~~~
~~~~~~~~~~~~~~~~~~~~~~~~~~~~~~~~~~~~~~~~~~~~~~~~~~~~~~~~~~~~~~~~~~~~~~~~~~~~~~~~~~~~(18)\\
\\
This is a relation of degree six representing six modes of wave propagation.

\begin{center}
\bf 6. CONDITIONS FOR INSTABILITIES\\
 \end{center}
We shall now analyse the wave instabilities under different conditions depending on direction of 
magnetic field.\\

{\bf 6.1   The magnetic field in the direction of wave propagation i.e.} 
${\bf \vec B_{0}\equiv (B_{0x},0,0)}$.
 The dispersion relation (18) reduces to ,\\
\\
$J(A^{2}-4\acute \omega ^{2}w_{0x}^{2})-4\omega \acute \omega A(w_{0y}^{2}+w_{0z}^{2})+JAV_{Ax}^{2}
(L_{y}^{-2}+L_{z}^{-2})+2\omega k\acute k^{2}(V_{Ax})^{2}(w_{0y}L_{z}^{-1}-w_{0z}L_{y}^{-1})
+2\omega \acute \omega \acute kV_{Ax}^{2}[L_{y}^{-1}(\omega w_{0z}-2iw_{0x}w_{0y})-L_{z}^{-1}
(\omega w_{0y}+2iw_{0x}w_{0z})]-\\
4\omega \acute \omega V_{Ax}^{2}(w_{0y}L_{y}^{-1}+w_{0z}L_{z}^{-1})^{2}=0$~~~
~~~~~~~~~~~~~~~~~~~~~~~~~~~~~~~~~~~~~~~~~~~~~~~~
~~~~~~~~~~~~~~~~~~~~~~~~~~~~~~~~~~~~~~~~~~~~~~~~~~~~~~~~~(19)\\
\\
{\bf where}  ${\bf\acute k=k-i/L_{x}, J=\omega ^{2}-k^{2}c_{s}^{2}+w_{g}^{2},V_{Ax}=B_{0x}/(4\pi \mu 
\rho _{0})^{1/2}}
=$ Alfven velocity,\\
${\bf A=\omega \acute \omega -k\acute kV_{Ax}^{2}, L_{s}}$ {\bf being the characteristic length 
of variation of magnetic field along s i.e. x, y and z direction for s=x,y,z.} 
It would be interesting to note that 2nd term shows that the waves will be affected by the joint 
action of non-uniformity of 
magnetic field along the wave propagation and rotation component perpendicular to wave propagation.
Furthar, the 4th, 5th and 6th term indicate that the waves which propagate along the magnetic field but 
perpendicular to 
non-uniformity of magnetic field (i.e. $L_{y},L_{z}= finite $) as well as rotation components 
(i.e $w_{0y}, w_{0z} \not= 0$),
must be affected by joint action of non-uniformity and rotation. Thus {\bf the joint action on the wave 
propagation along the magnetic 
field requires rotation component must be perpendicular to the wave propagation but non-uniformity may 
be parallel or 
perpendicular to wave propagation.}\\

{\bf Special case 1 :} We assume that the waves are propagating along the gradient of magnetic field 
(i.e. $ L_{x}$=finite but $ L_{y},L_{z}\longrightarrow \infty $ ) and rotation component is 
in the direction of wave propagation only i.e. $\vec w_{0}\equiv (w_{0x},0,0)$
(See Fig. 1) then from relation (19), \\
\\
$ J=0$~~~~~~~~~~~~~~~~~~~~~~~~~~~~~~~~~~~~~~~~~~~~~~~~~~~~~~~~~~~~~~~~~~~~~~~~~~~~~~~~
~~~~~~~~~~~~~~~~~~~~~~~~~~~~~(19.1) \\
 $\acute \omega (\omega \mp 2w_{0x})/k^{2}=V_{Ax}^{2}[1-i/(kL_{x})]$~~~~~~~~~~~~~~~~~~~~~~~~~~~~~~~~
~~~~~~~~~~~~~~~~~~~~~~~~~~~~~~~~~~~~~~~~~~~~~~~~~~~~~~~~~~~~~~(19.2)\\
\\
Now relation (19.1) and (19.2) yield [taking $\eta \longrightarrow 0$ i.e $\acute
 \omega \longrightarrow \omega $] the following,
\\
$\omega^{2}=k^{2}C_{s}^{2}-w_{g}^{2}$~~~~~~~~~~~~~~~~~~~~~~~~~~~~~~~~~~~~~~~~~~
~~~~~~~~~~~~~~~~~~~~~~~~~~~~~~~~~~~~~~~~~~~~~~~~~~~(19.3)\\
$k=\pm [m -n ^{2}]^{1/2}+in $~~~~~~~~~~~~~~~~~~~~~~~~~~~~~~~~~~~~~~~~~~~~~~~~~~~~~
~~~~~~~~~~~~~~~~~~~~~~~~~~~~~~~~~~~~~~~(19.4)\\
\\
where\\
 $m = \omega (\omega \mp 2w_{0x})/V_{Ax}^{2}$, $n =L_{x}^{-1}/2$\\

From (19.3) one can observe that the temporal stable mode of waves propagating along the 
common 
direction of magnetic field, its gradient and rotation requires: \\

Wave length , $\lambda < \frac{2 \pi C_{s}}{w_{g}} = \lambda_{J}$(Jean's wave length)

phase velocity, ($\frac{ \omega}{k}$) $= C_{s} [1-\frac{w_g^2}{k^2 C_s^2}]^{1/2}
< C_{s} = $ Sound Speed. ~~~~~~~~~~~~~~~~~~~~~~~~~~~~~~~~~~~~~~~~~(I a)\\

The second relation (19.4) shows that the waves can avoid intsability provided its
wave length  $\lambda << 4 \pi L_{x}$\\           
Otherwise, such waves will undergo damping in the direction of increasing magnetic
field  ($n>0$).
This instability will be further enhanced at least for one mode whose\\
   wave frequency( $\omega$ )$<$ 2 rotational frequency ($w_{0x}$  ).
 In this situation the Instabilty Factor is given by--\\
 
 $\beta=[\pm \frac {\omega(\pm 2w_{0x}-\omega)}{V_{Ax}^{2}}+(\frac {L_{x}^{-1}}{2})^{2}]^{1/2}+\frac {L_{x}^{-1}}{2}$
~~~~~~~~~~~~~~~~~~~~~~~~~~~~~~~~~~~~~~~~~~~~~~~~~~~~~~~~~~~~~~(I b)\\

All these exhibit that the presence of magnetic field and rotation in the direction of wave propagation 
imposes restrictions
on the stable and unstable mode's frequency and wave length.\\

{\bf Special case 2 :}  In this case we assume the magnetic field acts along the direction of wave 
propagation, 
but its gradient and rotation act perpendicular to the 
wave propagation direction i.e. $\vec B_{0}\equiv (B_{0x},0,0), L_{x},L_{y}\longrightarrow \infty , 
\vec w_{0}\equiv (0,0,w_{0z}) $ (See, Fig. 2) then relation (19) reduces to,\\
\\
$\omega \acute \omega +V_{Ax}^{2}(L_{z}^{-2}-k^{2})=0$~~~~~~~~~~~~~~
~~~~~~~~~~~~~~~~~~~~~~~~~~~~~~
~~~~~~~~~~~~~~~~~~~~~~~~~~~~~~~~~~~~~~~~~~~~~(20.1)\\
$J(\omega \acute \omega -k^{2}V_{Ax}^{2})-4\omega \acute \omega w_{0z}^{2}=0$~~~~~~~~~~~~~~~~~~~~~~~~~~~
~~~~~~~~~~~~~~~~~~~~~~~~~~~~~~~~~~~~~~~~~~~~~~~~~~(20.2)\\
\\
We may note that the 1st relation involves non-uniformity of magnetic field but is independent of rotation while the 
2nd relation involves rotation only but is independent of non-uniformity of magnetic field.
Now, the relation (20.1) yields (taking $ \eta \longrightarrow 0$ i.e. 
$\acute \omega \longrightarrow \omega $) ,\\
\\
$~~~~~~~~~\omega^{2}=V_{Ax}^{2} (k^{2}-L_{z}^{-2})$\\
Or,~~ $k^{2}=(\omega ^{2}+V_{Ax}^{2}L_{z}^{-2})/V_{Ax}^{2} > 0~~~~~~~~~~~~~~~~~~~~~~~~~~~~~~~~~~~
~~~~~~~~~~~~~~~~~~~~~~~~~~~~~~~~~~~~~~~$(20.3)\\
\\
These indicate that one mode will be temporally stable for

 $\lambda < 2\pi  L_{z}$.\\
The other mode is governed by (20.2) which, in turn, can be written as,\\
\\
$~~~~~~~~~\omega^{4}+\omega^{2}[-k^{2}C_{s}^{2}+w_{g}^{2}-4w_{0z}^{2}-k^{2}V_{Ax}^{2}]+
k^{2}V_{Ax}^{2}[k^{2}C_{s}^{2}-w_{g}^{2}]=0$\\
Or, ~~$k^{4}(C_{s}^{2}V_{Ax}^{2})-k^{2}[\omega ^{2}C_{s}^{2}+V_{Ax}^{2}(\omega ^{2}+w_{g}^{2})]+\omega ^{2}
(\omega ^{2}+w_{g}^{2}-4w_{0z}^{2})=0$ ~~~~~~~~~~~~~~~~~~~~~~~(20.4)\\
\\
If $\omega_{1}^{2}$ and  $\omega_{2}^{2}$ are two roots of (20.4) then,\\
$\omega_{1}^{2} + \omega_{2}^{2}= k^{2}C_{s}^{2}-w_{g}^{2}+4w_{0z}^{2}+k^{2}V_{Ax}^{2}$\\
$\omega_{1}^{2} . \omega_{2}^{2}=k^{2}V_{Ax}^{2}[k^{2}C_{s}^{2}-w_{g}^{2}]$\\
\\
Evidently, two roots $\omega _{1}^{2}$ and $\omega _{2}^{2}$ are positive i.e the mode will be 
temporal stable if

$k^{2}C_{s}^{2}>w_{g}^{2}$ ~~ i.e ~~ $\lambda < \lambda _{J}$= Jean's wave length (See, Ferraro and Plumpton, 1966).\\
\\
If $k_{1}^{2}$ and  $k_{2}^{2}$ are two roots of (20.4) then,\\
$k_{1}^{2} + k_{2}^{2} =[\omega ^{2}C_{s}^{2}+V_{Ax}^{2}(\omega ^{2}+w_{g}^{2})]/ (C_{s}^{2}V_{Ax}^{2})$\\
$k_{1}^{2} . k_{2}^{2} = \omega ^{2}(\omega ^{2}+w_{g}^{2}-4w_{0z}^{2})/ (C_{s}^{2}V_{Ax}^{2})$\\
\\
Clearly the two roots $k_{1}^{2}$ and $k_{2}^{2}$ are positive i.e the mode will
be spatially stable provided

$\omega^{2}+w_{g}^{2} > 4w_{0z}^{2}$\\
\\
Thus the stable mode must have

$\lambda < 2 \pi L_{z}$ ,~~~ $\lambda < \lambda _{J}$ ~~or~~ $\omega^{2}+w_{g}^{2} > 4w_{0z}^{2}$
~~~~~~~~~~~~~~~~~~~~~~~~~~~~~~~~~~~~~~~~~~~~~~~~~~~~~~~~~~~~~~~~~~~~~~~~~~~~~~~~~~~(II)\\
\\
Note that the second condition is Jean's condition. Thus the stable waves will be 
of higher frequency and of lesser 
wave length. Here both rotation and non-uniform magnetic field has been found
to constrain the frequency and wave length but separately. \\

{\bf 6.2 Magnetic field perpendicular to wave propagation }
Here we will investigate the waves moving in the direction of the magnetic field but field gradient and rotation act perpenducular 
to the direction of wave propagation i.e. $\vec B_{0}\equiv (0,B_{0y},0)$ or $\vec B_{0}\equiv (0,0,B_{0z})$ .\\

{\bf 6.2.1} For the first situation the dispersion relation (18) reduces to,\\
\\
$J[\acute \omega ^{2}(\omega ^{2}-4w_{0x}^{2})+\acute \omega V_{Ay}^{2}L_{z}^{-1}
(\omega L_{z}^{-1}-2iw_{0x}L_{y}^{-1})]-4\acute \omega ^{2} \omega ^{2}(w_{0y}^{2}+w_{0z}^{2})\\
+2\omega kL_{y}^{-1}L_{z}^{-1}(V_{Ay}^{2})^{2}(w_{0z} L_{z}^{-1}+i\acute k w_{0x})
-4\omega \acute 
\omega w_{0z}L_{z}^{-1}V_{Ay}^{2}(w_{0y}L_{y}^{-1}+w_{0z}L_{z}^{-1})-\\ 
\omega \acute \omega V_{Ay}^{2}(\omega ^{2}-4w_{0x}^{2})\acute k^{2}+
2i\omega \acute \omega L_{x}^{-1}L_{y}^{-1}(\omega w_{0z}+2iw_{0x}w_{0y})-
8i\omega \acute \omega w_{0x}w_{0z} L_{z}^{-1}V_{Ay}^{2}\acute k=0 $~~~~~~~~~~~~~~~~~~(21)\\
\\
where $V_{Ay}=B_{0y}/(4\pi \mu \rho _{0})^{1/2}, L_{s}$ {\bf being the characteristic length 
of variation of magnetic field along s i.e. x, y and z direction for s=x,y,z.} \\
It is interesting to note that there are product terms representing combined effect of rotation and magnetic 
field gradients (for example, 2nd part of 1st and 5th term, the 3rd, 4th, 6th and last term).\\
If we analyse the relation (21) we can arrive at the conclusion that 
{\bf for gradient of magnetic field along the direction of wave propagation, the combined effect is 
peroduced  provided $w_{0x}$ 
is present. However, 
the gradient of magnetic field along the direction of 
magnetic field can not produce combined effect. Again if the gradient of 
magnetic field is in the z-direction, then the combined effect is present for $w_{0z}$
only.}\\

{\bf Special case 3 :} Here we investigate the waves propagating (x-direction) perpendicular to 
magnetic field 
$\vec B_{0}\equiv (0,B_{0y},0)$ but its gradient and rotation both act perpendicular to both magnetic field 
and wave propagation 
(i.e. z-direction). Therefore, $L_{x},L_{y}\longrightarrow \infty$~~~ , $\vec w_{0}\equiv (0,0,w_{0z})$ 
(See, Fig. 3) then the relation (21) reduces to,\\
\\
$(J-4  w_{0z}^{2})(\omega \acute \omega +V_{Ay}^{2}L_{z}^{-2})=\omega ^{2}k^{2}V_{Ay}^{2}$
~~~~~~~~~~~~~~~~~~~~~~~~~~~~~~~~~~~~~~~~~~~~~~~~~~~~~~~~~~~~~~~~~(21.1)\\
\\
This relation reveals that {\bf the combined effect could be produced by magnetic field gradient and rotation 
provided both act perpendicular to magnetic field as well as wave propagaton}.\\
 Now, relation (21.1) yields (taking 
$\eta \longrightarrow 0$),\\
\\
$~~~~~~~w^{4} + w^{2}[V_{Ay}^{2}(L_{z}^{-2}-k^{2})-(k^{2}C_{s}^{2}-w_{g}^{2}+4w_{0z}^{2})]
+V_{Ay}^{2}L_{z}^{-2}[-k^{2}C_{s}^{2}+w_{g}^{2}-4w_{0z}^{2}]=0$\\
Or ~~ $k^{2}=(\omega ^{2}+V_{Ay}^{2}L_{z}^{-2})(\omega ^{2}+w_{g}^{2}-4w_{0z}^{2})/[\omega ^{2} V_{Ay}^{2}+
C_{s}^{2}(\omega ^{2}+V_{Ay}^{2}L_{z}^{-2})]$~~~~~~~~~~~~~~~~~~~~~~~~~~~~~~~~~~~~(21.2)\\
\\
If $\omega_{1}^{2}$ and $\omega_{2}^{2}$ are two roots of the first relation in (21.2) then,\\
$\omega_{1}^{2} + \omega_{2}^{2}= V_{Ay}^{2}(k^{2}-L_{z}^{-2})+ k^{2}C_{s}^{2}-w_{g}^{2}+4w_{0z}^{2}$\\
$~~~~~~~~~~~=k^{2}(C_{s}^{2}+V_{Ay}^{2})-(V_{Ay}^{2}L_{z}^{-2}+w_{g}^{2}-4w_{0z}^{2})$\\
$\omega_{1}^{2} . \omega_{2}^{2}=- V_{Ay}^{2}L_{z}^{-2}(k^{2}C_{s}^{2}-w_{g}^{2}+4w_{0z}^{2})$\\
\\
It is to be noted that if ~~$k^{2}C_{s}^{2} > w_{g}^{2}$  i.e. $\lambda < \lambda _{J}$,
 then both the roots can not be positive. 
This Jean's criteria can not ensure stable mode so long as magnetic field, its gradient and wave propagation 
 are mutually perpendicular to each other (See, Fig. 3). However,
 if the non-uniformity of magnetic field disappears i.e. 
 $L_{z}\longrightarrow \infty$, then the first relation of (21.2) yields, \\
 \\
 $\omega^{2}=k^{2}(V_{Ay}^{2}+C_{s}^{2})-w_{g}^{2}+4w_{0z}^{2}$\\
 \\
 This shows that for temporal stability Jean's criteron is sufficient but for spatial stability the requirement 
 is ~ $\omega^{2}+w_{g}^{2}>4w_{0z}^{2}$\\
Clearly, the temporal stability of the waves require that both the roots ($\omega_{1}^{2}$  and  $\omega_{2}^{2}$)
 must be positive. This means \\
 \\
$k^{2}(C_{s}^{2}+V_{Ay}^{2})>(V_{Ay}^{2}L_{z}^{-2}+w_{g}^{2}-4w_{0z}^{2})$ ~~~i.e ~~~
$\lambda < 2 \pi [\frac{C_{s}^{2}+V_{Ay}^{2}}{V_{Ay}^{2}L_{z}^{-2}+w_{g}^{2}-4w_{0z}^{2}}]^{1/2}$\\
and\\
 $k^{2}C_{s}^{2}-w_{g}^{2}+4w_{0z}^{2} < 0$~~~ i.e~~~ $\lambda >  2\pi \frac{ C_{s}}{[w_{g}^{2}-4w_{0z}^{2}]^{1/2}}$\\
 \\
These give a range of wave length for the temporal stability of waves. 
Namely,\\
 $ 2 \pi [\frac{C_{s}^{2}+V_{Ay}^{2}}{V_{Ay}^{2}L_{z}^{-2}+w_{g}^{2}-4w_{0z}^{2}}]^{1/2} >
 \lambda > 2\pi \frac{ C_{s}}{[w_{g}^{2}-4w_{0z}^{2}]^{1/2}}$~~~~~~~~~~~~~~~~~~~~~~~~~
~~~~~~~~~~~~~~~~~~~~~~~~~~~~~~~(III)\\
\\
Evidently, the rotation as well as magnetic field gradient both are capable to impose
restrictions on the upper bound of the wave length of temporally stable modes while only the rotation is 
capable to restrict the lower bound of the temporally stable mode.\\

{\bf  6.2.2 } For the second situation, the dispersion relation (18) reduces to ,\\
 \\
 $J[\acute \omega ^{2}(\omega ^{2}-4w_{0x}^{2})+\acute \omega V_{Az}^{2}L_{y}^{-1}
(\omega L_{y}^{-1}+2iw_{0x}L_{z}^{-1})]-4\acute \omega ^{2} \omega ^{2}(w_{0y}^{2}+w_{0z}^{2})\\
-2\omega kL_{y}^{-1}L_{z}^{-1}(V_{Az}^{2})^{2}(w_{0y} L_{y}^{-1}+i\acute k w_{0x})-4\omega \acute 
\omega w_{0y}L_{y}^{-1}V_{Az}^{2}(w_{0y}L_{y}^{-1}+w_{0z}L_{z}^{-1})-\\ 
\omega \acute \omega V_{Az}^{2}(\omega ^{2}-4w_{0x}^{2}) \acute k^{2}-
2i\omega \acute \omega L_{x}^{-1}L_{z}^{-1}(\omega w_{0y}-2iw_{0x}w_{0z})-
8i\omega \acute \omega w_{0x}w_{0y} L_{y}^{-1}V_{Az}^{2}\acute k=0 $~~~~~~~~~~~~~~(21.3)\\
\\
where $V_{Az}=B_{0z}/(4\pi \mu \rho _{0})^{1/2}, L_{s}$ {\bf being the characteristic length 
of variation of magnetic field along s i.e. x, y and z direction for s=x,y,z.}\\
The conclusions which can be drawn from the relation (21.3)are similar to that from (21).\\

{\bf \em SUMMERY :} We have discussed a few instabilities of the many cases which could be 
covered by the dispersion relation 
(18). For such cases pertaining to different situations we can study the instabilities, the role 
of combined effect of rotation 
and non-uniformity of magnetic field. But these 
will be done elsewhere. At present we can summerise the above study as follows.

When the wave propagates along x-direction we have considered three cases results of which are the following. \\
\vskip 0.2 cm 
\begin{tabular}{|c|c|c|c|c|c|}
\hline
{\bf Special} & ${\bf B_{0} }$ & ${\bf \nabla B_{0} }$ & $ {\bf w_{0} }$ & {\bf Dispersion Relation }&{\bf Figures}\\
{\bf Cases}& & & &  & \\
& & & & {\bf * Stability Condition} &\\
\hline 
 & & & & $ J=0$ &\\
1 & $B_{0x}$ & $\nabla _{x} B_{0x}$ & $w_{0x}$ & $\acute \omega (\omega \mp 2w_{0x})/k^{2}=V_{Ax}^{2}[1-i/(kL_{x})]$& Fig. 1\\
& & & & &\\
& & & & * $\lambda < \lambda_{J}$  or  $\lambda << 4 \pi L_{x}$ &\\
\hline 
& & & & $\omega \acute \omega +V_{Ax}^{2}(L_{z}^{-2}-k^{2})=0$ & \\
& & & & $J(\omega \acute \omega -k^{2}V_{Ax}^{2})-4\omega \acute \omega w_{0z}^{2}=0$ & \\
2 & $B_{0x}$ & $\nabla _{z} B_{0x}$ & $w_{0z}$ & & Fig. 2\\
& & & & * $\omega^{2}+w_{g}^{2}>4w_{0z}^{2}$ & \\
& & & & * or $\lambda < 2 \pi L_{z}$ or $\lambda < \lambda _{J}$ &\\
\hline
& & & & $(J-4  w_{0z}^{2})(\omega \acute \omega +
V_{Ay}^{2}L_{z}^{-2})=\omega ^{2}k^{2}V_{Ay}^{2}$ & \\
& & & & & \\
3& $B_{0y}$ & $\nabla _{z} B_{0y}$ & $w_{0z}$ & * $\omega^{2}+w_{g}^{2}>4w_{0z}^{2}$ & Fig. 3\\
& & & & * or $ 2 \pi [\frac{C_{s}^{2}+V_{Ay}^{2}}{V_{Ay}^{2}L_{z}^{-2}+w_{g}^{2}-4w_{0z}^{2}}]^{1/2}
 > \lambda > 2\pi \frac{ C_{s}}{[w_{g}^{2}-4w_{0z}^{2}]^{1/2}}$&\\
\hline 
\end{tabular}
\vskip 0.5 cm 
It is found that Jean's criterion is necessary but not sufficient for the temporal stability of waves in Special Cases 
1 and 2. However, the waves with shorter length and higher frequency are generally stable. For the Spatial stability 
the condition for Cases 2 and 3 are same. The Temporal stability for case 3 is too restricted to exist so 
long as non-uniformity of magnetic field exits.

\begin{center}
\bf 7. CENTRAL REGION OF GALAXIES\\
 \end{center}
The theory which has been discussed so far may be useful to explain different astrophysical phenomena 
like heating of the medium and mass outflow from it, provided, the medium could be treated 
essentially as MHD or Plasma.\\
 Let us first note down the characteristic features of the medium embeded in the central region 
of galaxies (See, e.g. Chakraborty and Bondyopadhaya,1998).\\
\\
{\bf1. Fluid  characteristics :} \\
We know that the ionised gas medium could be treated as fluid if the mean free path of 
the charged particles 
is less than the linear dimension of the medium ie. $\lambda _{m}<L$ (See, e.g. Woltjer, 1965).
 Now observations have revealed that huge material in different galactic central regions 
exists which
may satisfy this criterion (See, e.g. Krotik et al., 1981; Perola et al., 1986;
 Uberoi, 1988). 
We shall study the dynamics of this fluid in the central regions by the help of hydromagnetic
theory developed in the text (See also, Chakraborty and Bondyopadhaya, 1998). \\
\\
{\bf 2. Magnetic field:} \\
Woltjer (1965, 1971, 1990) discussed the important role of magnetic field in the activities 
of Galactic Nuclei,
Spiral Arms, Seyfert Galaxies and Radio jets (See also, Zeilik and Niley, 1997). This view 
was supported by 
many others (e.g. Osterbrock and Mathews, 1986). Infact, the existence of $\sim $ 1 gauss 
magnetic field in 
AGN has been reported by many authors (See, e.g. Rees, 1987, Wielebinski, 1990, Morris, 1990). 
{\bf There is, however, indication that the field is generally parallel to the plane 
in denser region (core) and perpendicular to the plane in  low density region (periphery)}
(See, Chuss et al., 2003).
The magnetic field parallel or perpendicular to the galactic plane is likely to be 
non-uniform.\\
\\
{\bf 3. Rotation:} \\
Generally rotation axis is either parallel or slightly inclined to the normal of the galactic 
plane. Therefore, for convenience we take the dominant component of rotation as perpendicular 
to the plane.\\
\\
{\bf 4. Division of Central Region:} \\
There are two zones in the central region of the galaxy.

{\bf i) Core Region:}  Obviously the core of the central region is a denser medium. Following 
Chuss et al., 2003, 
let us assume that this region has 
magnetic field parallel to galactic plane. As per discussion above the rotation can be 
taken as perpendicular 
to this galactic plane. Furthur, the magnetic field gradient can be taken parallel and 
perpendicular to the galactic plane.\\
{\bf ii) Peripheral Region:} Similarly, the peripheral region of the centre is a lighter 
medium. As per suggession 
Chuss et al., 2003, the magnetic field can be taken as perpendicular to the galactic 
plane. The magnetic field 
gradient and rotation can also be taken as before.

\begin{center}
\bf 8. STABILITY OR INSTABILITY OF THE CENTRAL REGION AND SIMILAR MEDIA
\end{center}
Now, we shall study the stability or instabilitiy of the wave propagation 
in the Core and Periphery of central region of the Galaxy.\\
 
{\bf i) The Core region:} It is of higher density where the magnetic field
 may be taken as parallel but its gradient may be
taken as perpendicular to the galactic plane. The MHD wave propagation in such 
system has been discussed in Special Case 3. The combined effect of 
rotation and magnetic field gradient has been found on the waves 
propagating parallel to the galactic plane. In this case, 
it is observed that the waves will be spatially stable for
 
 $4w_{0z}^{2} < \omega ^{2}+ w_{g}^{2}$.\\
And, the wave length of temporally stable wave requires the constrain (III) to
 be satisfied.\\
We may note that if L.H.S.is greater than R.H.S. then

 $w_{g}^{2}>4w_{0z}^{2}+C_{s}^{2}/L_{z}^{2}$~~ i.e. the density 
$\rho_{0}>\frac{[C_{s}^{2}/L_{z}^{2}+4w_{0z}^{2}]^{1/2}}{4\pi G}$\\
 The condition (III) is true when the galactic plane flow (x-direction)
 is perpendicular to magnetic field (y-direction). However, the flow along
 the magnetic field could be described by the theory outlined in Special
 Case 2 (Fig. 2). For stability such wave requires 

$\omega^{2}+w_{g}^{2}>4w_{0z}^{2}$ 
(as before) and $\lambda < \lambda_{J}$, $\lambda<2\pi L_{z}$\\

 When all these stable waves parallel to the galactic plane, and
 both along and perpendicular to the magnetic field continue to propagate
 for million of years those may feed the Spiral Arms or even may
 lead to the formation of Spiral Arms or other outer structures of the
 galaxies including our Galaxy. \\
  
{\bf ii)Peripheral Region :}
 This region is of low density  where the magnetic 
field, its gradient and rotation all act perpendicular to the galactic 
plane (as shown in Special Case 1, Fig. 1).
Here, the waves propagating perpendicular to the galactic plane
undergo instability due to the joint action of rotation and magnetic 
field gradient. One temporal stable mode can propagate whose phase velocity
    is less than sound velocity and whose wavelength $\lambda < \lambda _{J}$.
 The other mode may be stable provided $\lambda << 4 \pi L_{x}$. 
If the wavelength does not satisfy this condition the waves will be amplified
in the direction of decreasing magnetic field . In fact, the magnetic field decreases away
 from the galactic plane. In addition the waves with frequency
 $\omega < 2 w_{0x}$
will undergo instability with instability factor (I b) (as shown in Special Case 1), 
otherwise, I.F.=$1/(2L_{x})$ 
for the waves with period $\tau >\pi/w_{0x}$
these unstable waves can create a diffiused condition of material similar
  to Halo. While the stable waves can propagate in upward direction.
  However, it is interesting to note that if there is any flow towards 
the galactic plane in the direction of
  increasing magnetic field, the waves will be damped leading to the 
  heating of the medium. In the long run these may help to create hot Halo like 
  Solar Corona.
   
Now it is reported that from the galactic central region most of the 
energy come out in the form of infrared emission. In general, 
active galaxies have not only radio, u-v and x-ray but also infrared 
emission. In fact, later emission may be from heated dust or 
hot ionised gases (See, e.g. Zeilik, 1997). Heated dust around the 
nucleus of galaxies may absorb high energy radiation
emitted from energised Core and re-emits it at larger infrared wave 
length (See, e.g. Moche, 2004).\\

Infact, any instability may initiate heating of the medium via damping of
 MHD waves. Of course, 
not all waves are allowed to do it. Generally, waves with 
shorter wave lengths can pass through and longer waves are damped or
amplified ( See e.g. Sarkar and Bondyopadhaya, 2007).
Again, the system rotating with very high velocity may cause the 
instability of the MHD waves (See, Bondyopadhaya, 1972).
Therefore, both rotation and non-uniform magnetic field may be held
 responsible for the heating of the Central Region. 
But both in the Core and Periphery of the central region the waves with 
shorter wavelength or high frequency can move smoothly.
For such waves moving parallel or perpendicular to the galactic plane, when continues 
for long time 
may be responsible for the formation of the Spiral Arms or 
Halo region respectively.\\

We have seen in the text that the non-uniform magnetic field
when coupled with rotation is capable to 
produce instability. This type of instability could not be produced by
 them separately 
(See, e.g. Barnes, 1967; Bondyopadhaya, 1972; 
Paul and Bondyopadhaya, 1973; Yusef-Zadeh et al., 1986; Chakraborty and 
Bondyopadhaya, 1998). That is to say, this instability 
is a product function of non-uniform magnetic field and rotation. In 
reality, many astrophysical media posses both 
rotation and non-uniform magnetic field (the rotation axis, however,
 may not coincide with magnetic axis always).
Therefore, the theory like above may have good relevance for such media 
like AGN, Galactic Central Regions, 
Seyfert Galaxies and Infrared Galaxies. Hence,the joint effort of rotation 
and non-uniform magnetic field for initiating instability can be found there.
 In other words, the events occuring there 
may be at least partially due to joint action of non-uniform magnetic field 
and rotation.

\begin{center}
{\bf 9. NUMERICAL ESTIMATION}
\end{center}
In the previous section we have discussed the instability criteria
in the central region of galaxy (C.R.G.). Now, we will make a few numerical estimations 
of those criteria.
We have considered the central region of our galaxy as a hot plasma bed described by the MHD
equations (1 to 5) of Sec. 2 (See e.g., Chakraborty and Bondyopadhaya, 1993, 1998; Sarkar and 
Bondyopadhaya, 2007).
Let us first consider some physical parameters in the central region of our Galaxy.\\

The radius of the central region may be taken as 0.5 to 100 pc because well defined 
circumnuclear
gas disks have been found upto that region in a number of galaxies (Morris, 1998).\\

Further,the central region may be considered as a medium of uniform density, which is given by
 $\rho_{0}\approx 3.5 \times 10^{-20} g$ $cm^{-3}$ 
i.e.  $10^{36} g$ $pc^{-3}$ (See Balick and Sanders, 1974; Genzel et al., 1985; Lo, 1986 ).
Consequently, gravitational frequency becomes
 $w_{g}=(4 \pi G \rho_{0})^{1/2}\approx 1.7\times 10^{-13} s^{-1}\approx 5.36 $ per mil. yr.\\
 
The adiabatic sound speed is given by $C_{s}\approx 6.5 pc $ per mil yr (for $T_{0}=10^{4} K$ 
at 10 pc from the centre, vide Yusef-Zadeh et al., 1984).\\

The strength of general magnetic field is 4-6 $\mu G$ with a cell size of 10-100 pc 
(See, Ohno and Shibata, 1993).
Let us take magnetic field strength as $5 \mu G$ and the variation of magnetic field as 
$2 \mu G$
 over 50 pc range. Thus the characteristic length of variation $L_{x}$ 
or $L_{z}\approx 25 pc $. Hence,
 Alfven velocity $V_{A}\approx 1.4\times 10^{-18}pc$ $s^{-1} 
\approx 4.41 \times 10^{-5}$ pc per mil. yr .\\

The central region of the Galaxy rotates like a rigid body. Beyond this region the 
velocity first
decreases and then increases. At nearly 8 kpc from the centre the velocity becomes maximum. The
Sun is situated at 8.5 kpc from the centre having linear velocity $220 km s^{-1}$. Then Sun's
angular velocity about galactic centre $ \approx 84.6\times 10^{-17} s^{-1}$. 
(See, Kartunnen, 2007). Let us take this value as same as the angular velocity 
of central region i.e.
 $w_{0z}\approx 84.6\times 10^{-17} s^{-1} \approx 2.7 \times 10^{-2}$ per mil. yr.\\
 
Therefore, we obtain a
 critical wave length (i.e. Jeans wave length)

 $\lambda_{J}= 2 \pi C_{s}/w_{g} \approx 7.62 pc$.
 
Next, let us find what are the waves which can propagate or which will suffer instability inside the two areas of the 
central region of our Galaxy.\\

{\bf i) Peripheral Region :} (Special Case 1, Fig. 1)

Here the magnetic field, its gradient and rotation all act perpendicular to the galactic plane. One modes 
of the waves moving perpendicular 
to this plane will be temporal stable provided 

$\lambda < \lambda_{J}\approx $7.6 pc and the other mode require

 $ \lambda <<4\pi L_{x}\approx $ 315 pc. Conversely 
if the waves have length greater than 7.6 pc those will suffer instability. In addition, if $\lambda$ 
is not much less than 315 pc
some more modes will be unstable. Moreover, the waves having frequency less than $2w_{0x}\approx  17\times 10^{-16} s^{-1}$ 
(i.e. $\tau > 10^{8}$yr) the spatial instability will be enhanced. The actual instability factor 
is given by $\beta $ in Special case 1.\\

{\bf Such waves will be amplified leading to the defused condition if it moves along the decreasing direction 
of magnetic field i.e. away from the Galactic plane towords
Halo. If, however, there is inflow of MHD waves from outside towards the Galactic plane i.e. in the direction 
of increasing magnetic field
then the waves will be damped leading to the heating of the Halo  \footnote{We make a note of the 
fact that the source of energy of the solar corona is still not known but it is thought to be some how associated with 
magnetic reconnection and, therefore must be understood in terms of MHD process (See, Cravens, 1997).} }.\\

{\bf ii) Core Region :} (Special case 2,3. Fig 2,3)

The spatial stability of waves in this region require $w_{g}> 2w_{0z}$. As we have already discussed here 
$w_{g}\simeq 1.7\times 10 ^{-13}
 s^{-1}$ and $w_{0z}\simeq 8.5 \times  10 ^{-16} s^{-1}$. Thus the inequality is satisfied. 
Hence, waves of all frequencies will spatially stable.

Now, the waves will propagate, in the direction perpendicular to the magnetic field, smoothly (i.e. temporally stable) 
must have the wave length restricted by the following inequality 

 $  2 \pi [\frac{C_{s}^{2}+V_{Ay}^{2}}{V_{Ay}^{2}L_{z}^{-2}+w_{g}^{2}-4w_{0z}^{2}}]^{1/2} >
\lambda > 2\pi \frac{C_{s}}{[w_{g}^{2}-4w_{0z}^{2}]^{1/2}}$\\

Incidentally, $w_{g} (\approx 1.7\times 10^{-13}) >> 2w_{0z} (\approx 1.7 \times 10^{-15})$ and\\
 $C_{s} >> V_{Ay}$, $V_{Ay}/L_{z}\approx 3\times 10^{-20}s^{-1}$. These values will make R.H.S and L.H.S 
of the above inequality almost same. Therefore, no temporally stable wave can exist. However, those waves 
moving along the magnetic field will be temporally stable whose wave length is given by 

$\lambda < 2 \pi L_{z}\approx 157.5 pc$.\\

{\bf As we have seen that in the central region of our Galaxy $w_{g} >> 2 w_{0z}$ and $C_{s}>>V_{Ay}$ 
but the situation might be different in the long past in the life of our Galaxy itself. 
When both magnetic field and rotation was much higher. Similar 
situation may be present in some other galaxies also. For example, in AGN magnetic field is $10^{6}$ 
times (See, Sec. 7) the field that we have taken, so that the contribution due to magnetic field or its 
gradient is much more significant. The derived condition could be applicable in those 
galaxies particularly at the time of formation.}

\begin{center}
{\bf 10. CONCLUSION}
\end{center}
In this paper the General Dispersion Relation (G.D.R.) for the unidirectional (x-direction) wave propagation 
has been derived in 
the rotating medium where magnetic field is 
non-uniform. Then the guiding dispersion relations have also been obtained for magnetic field either along 
or perpendicular to wave propagation. 
There could be many different situations depending on the magnetic field, its gradient and the direction of 
rotation. But out of these only 
three Special Cases have been discussed.

A number of temporal/spatial instability/stability conditions have been derived. Comparing the Special Cases 
it is observed that {\bf the combined effect of 
rotation and magnetic field gradient is present in the product form provided the wave propagates along the 
magnetic field. The coupled action of rotation and magntic field's non-uniformity have also been noticed 
when the wave propagates perpendicular to the magnetic field and its gradient.} In general 
the instability conditions reveal that the waves having shorter frequency or longer wave length become unstable.

The instability conditions, thus derived, could be used to study different phenomena occuring in the the medium 
similar to the central region of our Galaxy.
It is suggested that such stabilities or instabilities are more significant in the early stage of evolution of many 
galaxies when both rotation and 
magnetic field are much higher. Those waves which can propagate from the central region may help in the formation 
of Spiral Arms and Halo of the galaxy in long run.

\begin{center}
\bf REMARKS
\end{center}
i) Mathematical analysis have been given more stress in this paper.\\
ii) A number of sub cases have also been analysed but all those are not included in this paper.\\
iii) The theory of the formation of hot Halo and Spiral Arms from the Galactic central material appears 
to be consistence with the underlined proposition of this paper.\\
iv) It is well known that the electrical resistivity causes the wave unstable. If $\eta$ does not tend to 0, 
the dispersion relations obtained in the text could be easily derived. This will, however, be considered in separate 
communication. 

\newpage
\begin{center}
\bf ACKNOWLEDGEMENT 
\end{center} 
Authors are thankful to the Department of Mathematics, Jadavpur University, Kolkata-700032, India for extending cooperation 
to work. Authors are also thankful to Dr. S. N. Chakraborty of P.K.H.N. Mahavidyalaya, Kanpur, 
Howrah, W.B., India and Z. Sarkar of N.I.T., Kamarhati, 24 Parg(N), W.B., India 
for their encouragement and giving assistance in the preparation of this paper. 

\begin{center}
\bf REFERENCES 
\end{center} 
Balick, B. and Sanders, R.H., Astrophys. J., {\bf 192}, 325 (1974).\\
Barnes, A., Phys. Fluids, {\bf 9}, 1483 (1967).\\ 
Barnes, A., Ap.J., {\bf 154}, 751 (1968).\\
Blandford, R.D., in Active Galactic Nuclei, T.J.L, Courvoisier and M.Mayor,  p.161, 

Springer-Verlag (1990).\\
Bondyopadhaya, R., Indian J. Phys., {\bf 46}, 281 (1972).\\
Bondyopadhaya, R., Paul, S.N, Fijika, {\bf 5}, 139 (1973).\\
Bondyopadhaya, R., Indian J. Phys., {\bf 48}, 585 (1974).\\
Bondyopadhaya, R. et al., Astronomy J. (Russian), {\bf 55}, 4, Moscow (1978).\\
Chakraborty, S.N., Bondyopadhaya, R., Astrophys. Space Sci., {\bf 203}, 1 (1993).\\
Chakraborty, S.N., Bondyopadhaya, R., Bull. Astr. Soc. India, {\bf 26}, 625 (1998).\\
Chakraborty, S.N., in Ph.D Thesis (Jadavpur University, India) (1998).\\
Chuss, D.T. et al., Astrophysical J., {\bf 599}, 1116 (2003).\\
Cravens, T.E., in Physics of Solar System Plasmas, p. 173, Cambridge (1997).\\
Ferraro, V.C.A and Plumpton, C., in An Introduction to Magnetofluid
Mechanics,

 p. 116, Clarendon, Oxford (1966).\\
Genzel, R. et al., Astrophys. J., {\bf 25}, 377 (1985).\\
Gilman, P.A., Ap. J., {\bf 162}, 1019 (1970).\\
Gold, L., Phys. Rev., 137A,4A,A, 1083 (1965).\\
Kartunnen, H. et al., in Fundamental Astronomy, p.359, Springer (2007).\\
Krotik, J.H., Mc. Kee, C.F, Tarter, C.B, Ap. J., {\bf 249}, 422 (1981).\\
Lo, K.Y., Astron. Soc. Pacific,, {\bf 98}, 179 (1986).\\
Moche, D.L, in Astronomy, p.163, Wiley J. and Sons (2004).\\
Morris, M., IAU Symp, {\bf 140}, 361 (1990).\\
Morris, M., IAU Symp, {\bf 184},3 (1998).\\
Ohno, H., Shibata, S., Mon. Not. R. Astron. Soc. {\bf 262}, 953 (1993).\\
Osterbrock, K.D.E., Mathews, W.G., ARA and A,{\bf 24}, 171 (1986).\\
Perola, G, Cetal, Ap. J., {\bf 306}, 508 (1986).\\
Rees, M.J, MNRAS, {\bf 228}, 47 (1987).\\
Rosen, A. and Hardee, P.E., Ap. J., {\bf 452},750 (2000).\\
Sarkar, Z., Bondyopadhaya, R., Astr. Sp. Sc., { \bf 312}, 34, 215-225 (2007).\\
Sturrock, P, Phys. Rev., {\bf 117}, 1426 (1960).\\
Sturrock, P.A, Bull. Am. Phys. Soc. {\bf 11}, 766 (1966).\\
Sudan, R.N, Phys. FL, {\bf 8 }, 1899 (1965).\\
Uberoi, C., in Introduction to Unmagnetized Plasmas, Prentice Hall of India, 

New Delhi, India (1988).\\
Wielebinski, R., IAU. Symp, {\bf 140}, 535 (1990).\\
Wiita, Paul, J., in Cosmic Radio Jets, Frontiers in  Astrophysics, S.P. Sengupta (edi.).\\ 

 Allied (2002).\\
Woltjer, L., in Galactic Structure, p. 531, (edi.) A. Blaauw and M.Schmidt, the Univ. of 

Chicago (1965).\\
Woltjer, L., in Study Week on Nuclei of Galaxies, ed. D.J.K. O'Connell, p. 741, Amsterdam (1971).\\
Woltjer, L., in R. Beck., p. 533, P. P. Kronberg and R. Weilebinski(edi.), IAU Symp.,{\bf 140} (1990).\\	
Woltjer, L., in Active Galactic Nuclei, p. 1, (edi.) T.J.L. Courvoisier and M. Mayor, Springer-Verlag (1990).\\
Yusef - Zadeh, F., Morris, M. and Chance, D., Nature, {\bf 310}, 557 (1984).\\
Yusef - Zadeh, F., et al., Astrophys.J., {\bf 300}, L47 (1986).\\
Zeilik, M., in Astronomy, p. 460, Wiley J. and Sons (1997).\\

\newpage
\begin {figure} [h]
\centering 
\includegraphics{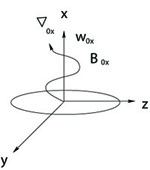}

\underline{FIG. 1}
\end{figure}
\begin {figure} [h]
\centering 
\includegraphics{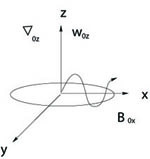}

\underline{FIG. 2}
\end{figure}
\begin {figure} [h]
\centering 
\includegraphics{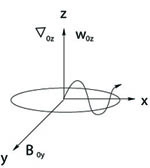}

\underline{FIG. 3}
\end{figure}
\end{document}